\documentclass[12pt]{article}
\usepackage{CJKutf8}
\usepackage{adjustbox}
\usepackage{amsmath}
\usepackage{amsthm}
\usepackage{amssymb}
\usepackage{graphicx}
\usepackage[utf8]{inputenc} % `utf8` option to match Editor encoding
\usepackage[english]{babel}
\usepackage[T1]{fontenc}
\usepackage{tikz}
\usetikzlibrary{cd}
\usetikzlibrary{arrows,automata}
\usetikzlibrary{babel}
\usepackage{mathtools}
\usepackage{float}%
\usepackage{bbm}
\usepackage{amsfonts}
\usepackage{xcolor}
\usepackage[most]{tcolorbox}
\usepackage{color}
\usepackage{enumerate}
\usepackage{microtype}
\usepackage{lmodern}
\usepackage{systeme}
\usepackage{makeidx}
\usepackage{hyperref}
\usepackage{pdflscape}
\usepackage{authblk}
\hypersetup{
	colorlinks=true,
	linktoc=all,    
	linkcolor=black,
}
\usetikzlibrary{cd}
\title{Introduction of L0 norm and application of  L1 and C1 norm in the study of time-series of indices of cryptocurrencies, South American currencies, banking indices and European indices.}
\author{Víctor Ujaldón García   }
\affil[]{\small{vujaldga7@alumnes.ub.edu  / +34 640272324 / victoru667@gmail.com}}

\theoremstyle{definition}

\theoremstyle{remark}

\theoremstyle{definition}

\usepackage{import}
\usepackage{xifthen}
\usepackage{pdfpages}
\usepackage{transparent}
\usepackage{svg}

\usepackage{pstricks}
\usepackage{pstricks-add}
\usepackage{epsfig}
\usepackage{pst-all} % For gradients
\usepackage{pst-plot} % For axes
\usepackage[space]{grffile} % For spaces in paths
\usepackage{etoolbox} % For spaces in paths
\makeatletter % For spaces in paths
\patchcmd\Gread@eps{\@inputcheck#1 }{\@inputcheck"#1"\relax}{}{}
\makeatother
\makeindex
\begin{document}
	
	\maketitle
	\begin{abstract}
		Four markets are considered: Cryptocurrencies / South American exchange rate / Spanish Banking indices and European Indices and studied using TDA (Topological Data Analysis)  tools. These tools are used to predict and showcase both strengths and weakness of the current TDA tools. In this paper a new tool $L0$ norm is defined and complemented with the already existing $C1$ norm.
	\end{abstract}
\section{Introduction}
To study market stability is to study abrupt changes on the market behavior. Time-series and statistical analysis have been found useful but taxing to compute; both from a theoretical and practical point of view. Topological Data Analysis (TDA) uses topological tools, mainly persistence homology, to predict critical transitions and the state of the market and future behavior. In fact TDA is capable of recognizing global patterns and group tendencies all of this without any requirements of expected volatility , bias, etc.  The ability to detect and predict big shifts and common group patters without an underlying distribution makes computation quick at the cost of specific predictions. That is: TDA can help establish group movements inside a market and if those movements will prevail but  not where to. In a sense the L0,L1 and C1 norms function similar to the $ R^2$ index.

\section{Informal introduction to  TDA}

Consider a point cloud $X$,  the Vietoris-Ripps complex $C_\epsilon(X)$ for a given $\epsilon>0$ consists of all the simplices (that is triangular polygons) of $X$ with diameter less or equal to $\epsilon$. Each $C_\epsilon(X)$ is a chain complex and as such we can consider a long exact sequence via the boundary operator which ,informally ,  takes edges to vertices. Let us consider the following example: set $X$ to be a point cloud of 3 elements $X = \{(0,0),(1,1), (1,0)\}$ then we have
\begin{align}
	C_0(X ) &= \{(0,0),(1,1), (1,0)\}\\
	C_1(X) &= \{[(0,0),(1,0)],[(1,0),(1,1)],(0,0),(1,1), (1,0)\} \\
	C_2(X) &= \{[(0,0),(1,0),(1,1)],[(0,0),(1,0)],[(1,0),(1,1)],[(0,0),(1,1)],(0,0),(1,1), (1,0)\}
\end{align}
where $[a,b]$ is the line segment between $a$ and $b$ and $[a,b,c]$ is a triangle with edges $a,b,c$. In this case $C_0(X)$ only has 3 vertices so the boundary operation does nothing. For $C_1$ we have 2 edges and 3 vertices the boundary operator would map $[(0,0),(1,0)]$ to $(1,0)-(0,0)$. In general the boundary operator is defined as:
\begin{align}
	\partial ([v_0,...,v_n]) = \sum_i (-1)^i [v_0,...,\hat{v_i},...,v_n]
\end{align}
It is clear that any closed $n-$dimensional path (in our example that would be $[(0,0),(1,0)]+[(1,0),(1,1)]+[(1,1),(0,0)]$) yields 0 when we apply the boundary operator. Studying the kernel  of the boundary operator, in particular the homology, allows us to determine holes in the data set for each given $\epsilon$. Persistence homology is the study of the preservation of $n$-dimensional holes in our data set $X$. For each $\epsilon$ we can determine if $C_\epsilon(X)$ has $n$-dimensional holes, if a new hole appears we establish $\epsilon$ as the birth of that hole, and if a hole disappears we establish $\epsilon$ as the dying value. Eventually we will have no holes left and we can represent the birth and death of $n$-dimensional holes in a matrix $\mathfrak{D}_n$. Finally we can compute the landscape function for each pair of birth and death $(b,d)$ in $\mathfrak{D}_n$ as 
\begin{align}
	f_{b,d}(x)= \left\{\begin{matrix}
		x-b & \textit{if } b<x\leq \frac{b+d}{2}\\
		d-x & \textit{if }  \frac{b+d}{2}<x\leq d\\
		0 & \textit{otherwise.}
	\end{matrix}\right.
\end{align}
The union of every $n$ dimensional hole landscape function is the persistence landscape function $\lambda^k(x)= kmax\{f_{b,d}(x)\}$. The more area a landscape function has the ''bigger'' the hole is. With this intuition in mind we consider the $Lp$ norm $||\Lambda||_{Lp} = \sum_k ||\lambda^k(x)||_p$ where $||f||_p := \left(\int f^p d\mu\right)^{1/p}$  denotes the standard $L^p$ norm for a given measure $\mu$.
\\
For a time series point cloud $\{X^a\}_{a\in (0,\mathfrak{T})}$ and a time window $T$ we define $||\Lambda_t||_{Lp} = \sum_k ||\lambda_t^k(x)||_p $ the $Lp$ norm at time $t$ of the point cloud $\{X^a\}_{a\in (t,t+T) }$.
Similarly we can define the $Cp$ norm as  $||\Lambda_t||_{Cp} = ||\Lambda_t||_{Lp}+||\Lambda_t||_{Lp}-||\Lambda_{t-1}||_{Lp}$ for $t\geq 1$. (More information on $Cp$ norm on \cite{c1}).

\section{Procedure}
Outside $ ||\lambda^1(x)||_1$ the rest don't yield much information for dimensions bigger than 1 and so from now on the $L1$ norm of the persistent landscape function  will be considered as the $L_1$ norm of $\lambda^1(x)$. Given the taxing computation times for higher homologies we will only consider the persistence matrix $\mathfrak{D}_0$ and  $\mathfrak{D}_1$. Finally the $L1$ norm  $||\lambda^1(x)||_1$ for $\mathfrak{D}_0$ is always the maximum and so we will consider the $L0$ norm as the $L_1$ norm of  $\lambda^2(x)$ for $\mathfrak{D}_0$. So from now on:
\begin{itemize}
	\item The $L1$ norm of $\{X^a\}_{a\in (0,\mathfrak{T})}$ is the area of the first landscape function for 1 dimensional holes over a time window $T$. Bigger $L1$ implies less stability.
	\item The $C1$ norm of $\{X^a\}_{a\in (0,\mathfrak{T})}$ is the corrected area of the first landscape function for 1 dimensional holes over a time window $T$. Bigger $C1$ implies critical states.
	\item The $L0$ norm of $\{X^a\}_{a\in (0,\mathfrak{T})}$ is the area of the  second landscape function for 0 dimensional holes over a time window $T$. Bigger $L0$ implies a dispersed market.
\end{itemize}
To construct a time series $\{X^a\}_{a\in (0,\mathfrak{T})}$ able to study market data we consider the closing values of  the stock/index/currency exchange value of $n$ units and compute the $log$ scale. That is: for a 1 dimensional data set  $\{x_{i,j}\}$ we consider $X_j = \{\log(x_{i+1,j}/x_{i,j})\}$ and the time value $\{X_j^a\} = \{\log(x_{i+1,j}/x_i,j)\}_{i = a}$. Finally we consider the temporal cloud point $\{X^a\}_{a\in (0,\mathfrak{T})} = \ \{X_1^a,...,X_n^a\}_{a\in (0,\mathfrak{T})}$ and we study the $L0,L1,C1$ norms with a time window $T$. In every case we used the standard time window $T=30$ and $j=4$ units.
\newpage
\section{Results}
\subsection{European stock market indices, why $L1$ matters}
Consider the following stock indexes of European markets Ibex35, BFX, ATX, CAC40(FCHI). The L1 norm computes the ''chaos'' in an informal sense of the market. Suppose a market with two indices $X,Y$ if both behave monotonically it's clear that $L1$ will be 0. However the more erratic both of them move the more ''holes'' the data will have and we will see an increase in the $L1$ norm. Erratic behavior is classically seen in unstable markets. As we see in \textbf{Figure 1} the L1 norm coincides with the 2008 market crash and if we look at the normalized relative value (that is the normalization of every market value separately) we can ''feel'' the erratic behavior around 2008-2009, also we see a spike around 2020 with the covid crisis but as it was a global shock (affecting every country almost at the same time) the $L1$ norm is not as big as the 2008 crisis. 
\begin{figure}[h]
	\centering
	\includegraphics[width=0.7\linewidth]{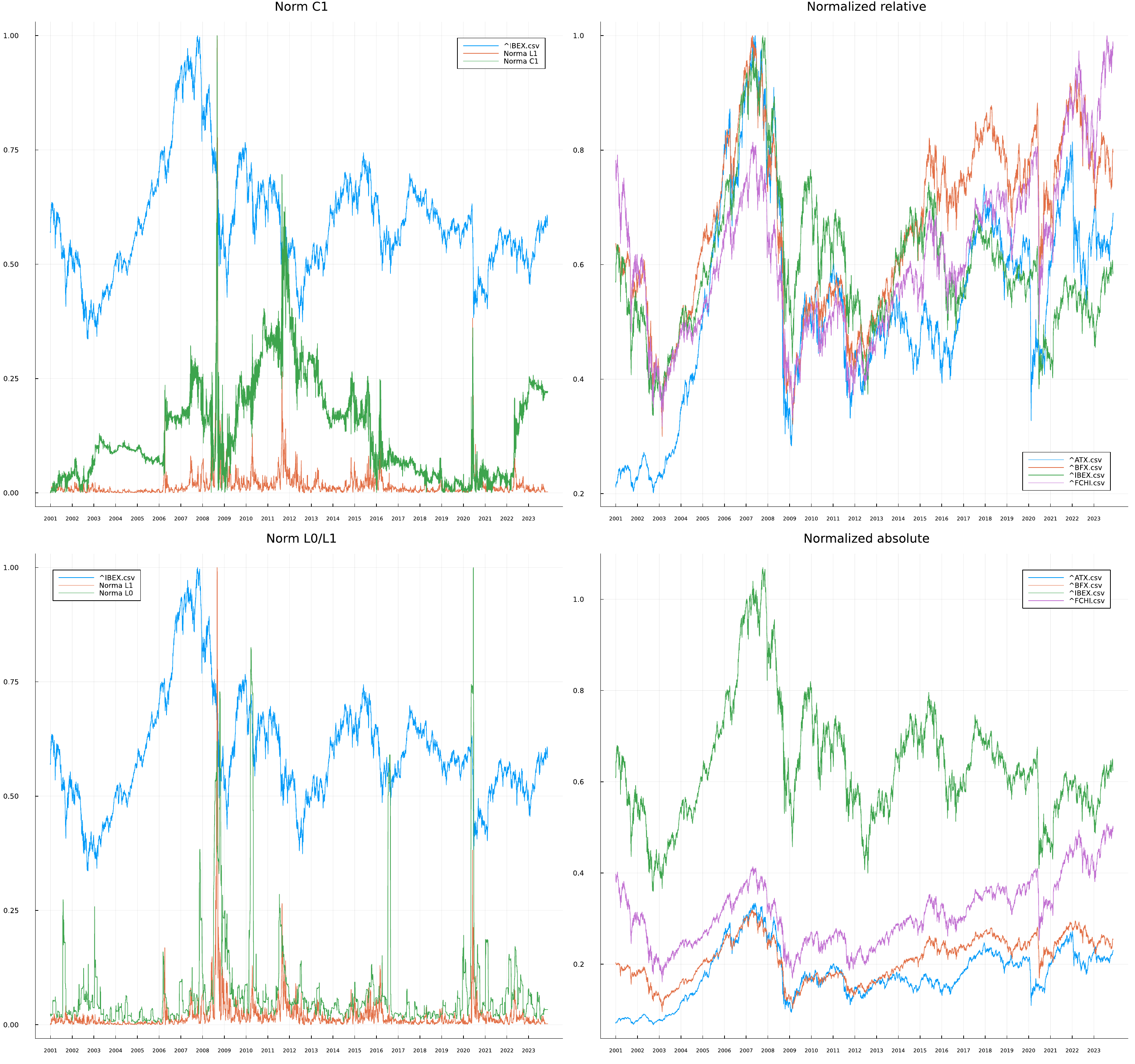}
	\caption{L0,L1,C1 and normalized comparison of european indices IBEX/ATX/BFX/FCHI }
	\label{fig:banco}
\end{figure}
\subsection{South American currencies, why $L0$ matters}
Consider now the currency change market of  USD to ARS (argentinian peso) / CLP ( chilean peso) / (PYG (paraguayan peso) / UYU (uruguayan peso) . For the past months Argentina has been suffering an incredible increase in their inflation and therefore the ratio of USD to ARS has skyrocketed. It's is cleary that the market formed by those indices is not ''stable'', 3 of them might be but clearly not ARS.  If every market was increasing at the same speed it would work as an homothety and therefore the topological structure would prevail (\cite{c2}). But given that one of the indices is increasing at such high rate the $L1$ norm stays 0 as any Ripps complex will be killed before obtaining meaningful data. The $L0$ norm determines a type of ''diameter'' of the data in a time lapse. We see that even though the $L1$ is almost 0 for the past months the $L0$ norm increases drastically showcasing the state of the market as one of the currencies suffers and exponential decrease in value. 
\begin{figure}[H]
	\centering
	\includegraphics[width=0.7\linewidth]{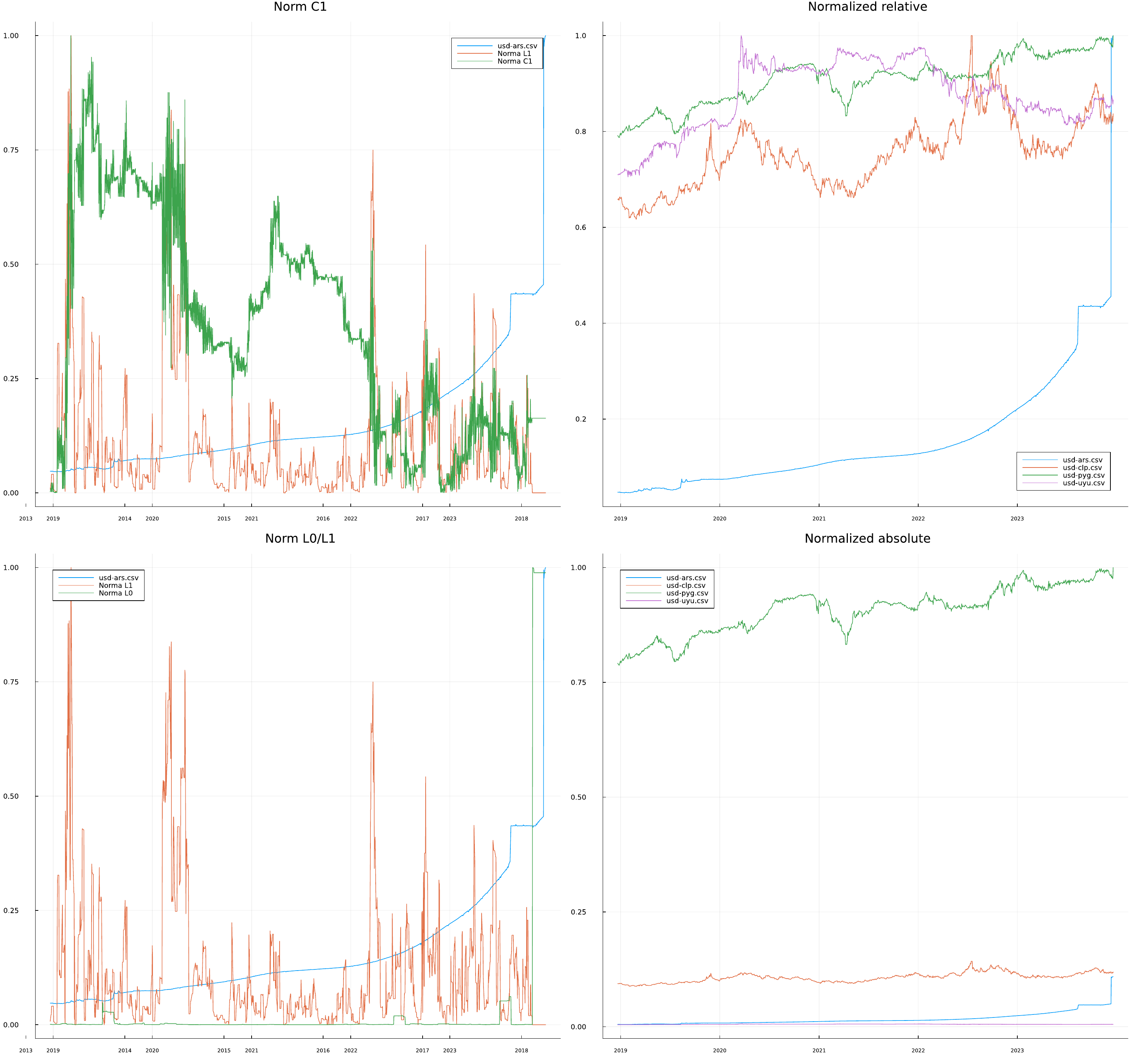}
	\caption{L0,L1,C1 and normalized comparison of South American currencies indices USD to ARS/CLP/PYG/UYU }
	\label{fig:banco}
\end{figure}

\subsection{Crypto currencies, why $C1$ matters}
Now we take a look at the cryptocurrency market given by  BTC/DOGE/XRP/ETH to USD. The $L0$ norm is fairly low therefore the indices are increasing /decreasing in a rather ''similar'' rate except from a booming in 2021 where we can see in the normalized relative graph that BTC exploded earlier than the others. The $L1$ norm outside this booming is also fairly stable around 0, this would lead to believe that the cryptocurrency market formed by those 4 changes is stable. However if every exchange is equally erratic then  $L1$ will not showcase the erratic behavior it will showcase that as a unit they move together. To compensate the $C1$ norm considers the distance '' jumped'' between times. Here the $C1$ norm stays really high showcasing that the cryptocurrency exchange market of those 4 indices is highly volatile and therefore in a critical transition state and with the low values of $L0,L1$ we can establish that the volatility is endemic , that is: those 4 indices are ''similarly chaotic''.
\begin{figure}[H]
	\centering
	\includegraphics[width=0.7\linewidth]{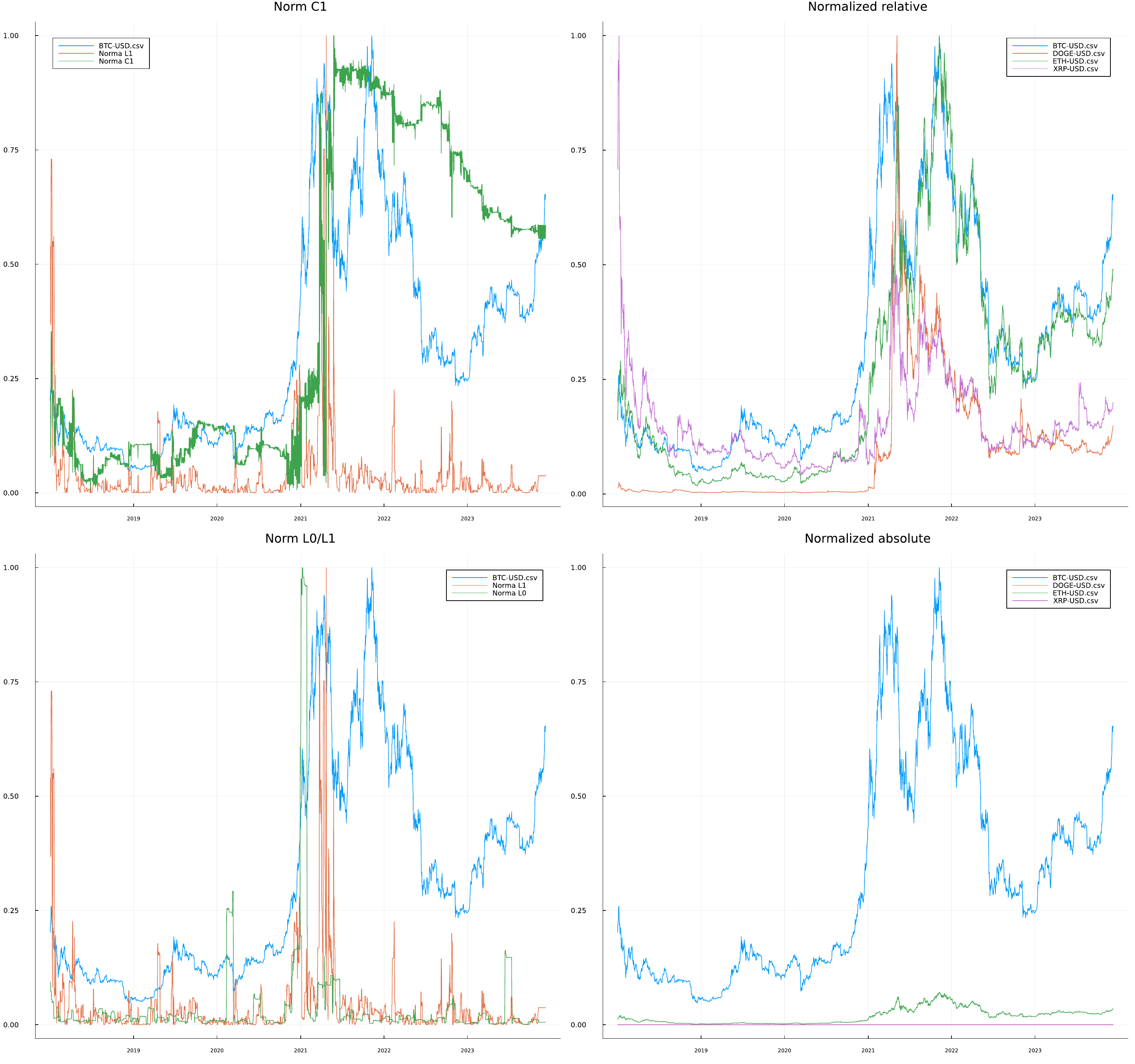}
	\caption{L0,L1,C1 and normalized comparison of cryptocurrency  indices BTC/DOGE/XRP/ETH to USD}
	\label{fig:banco}
\end{figure}

\subsection{Spanish Banks}
Finally we use everything to study the state of Spanish Banking via the banks: La Caixa, BBVA, Santander , Sabadell.
First of all by looking at the normalized relative values we can ''sense'' that every bank is moving in a similar direction (outside maybe 2020-2021 when Sabadell had a bit of a collapse) even though the absolute values are all very different with BBVA in the clear lead. Looking at $L0$  it's clear that those 4 banks increases are close together, the low  $L1$ establishes that the volatility of the market is  close; that is every bank has similar volatility. Finally $C1$ indicates that the global volatility of the market has been slowing and for the past year has been low; that is not anly we have a close market with similar volatility the volatility (or critical transition indicator) of those 4 indices is actually low. And so our initial feeling of closeness and stability of this market as each bank increases in almost the same way is validated via $L0,L1$ and $C1$ norms.
\begin{figure}[H]
	\centering
	\includegraphics[width=0.7\linewidth]{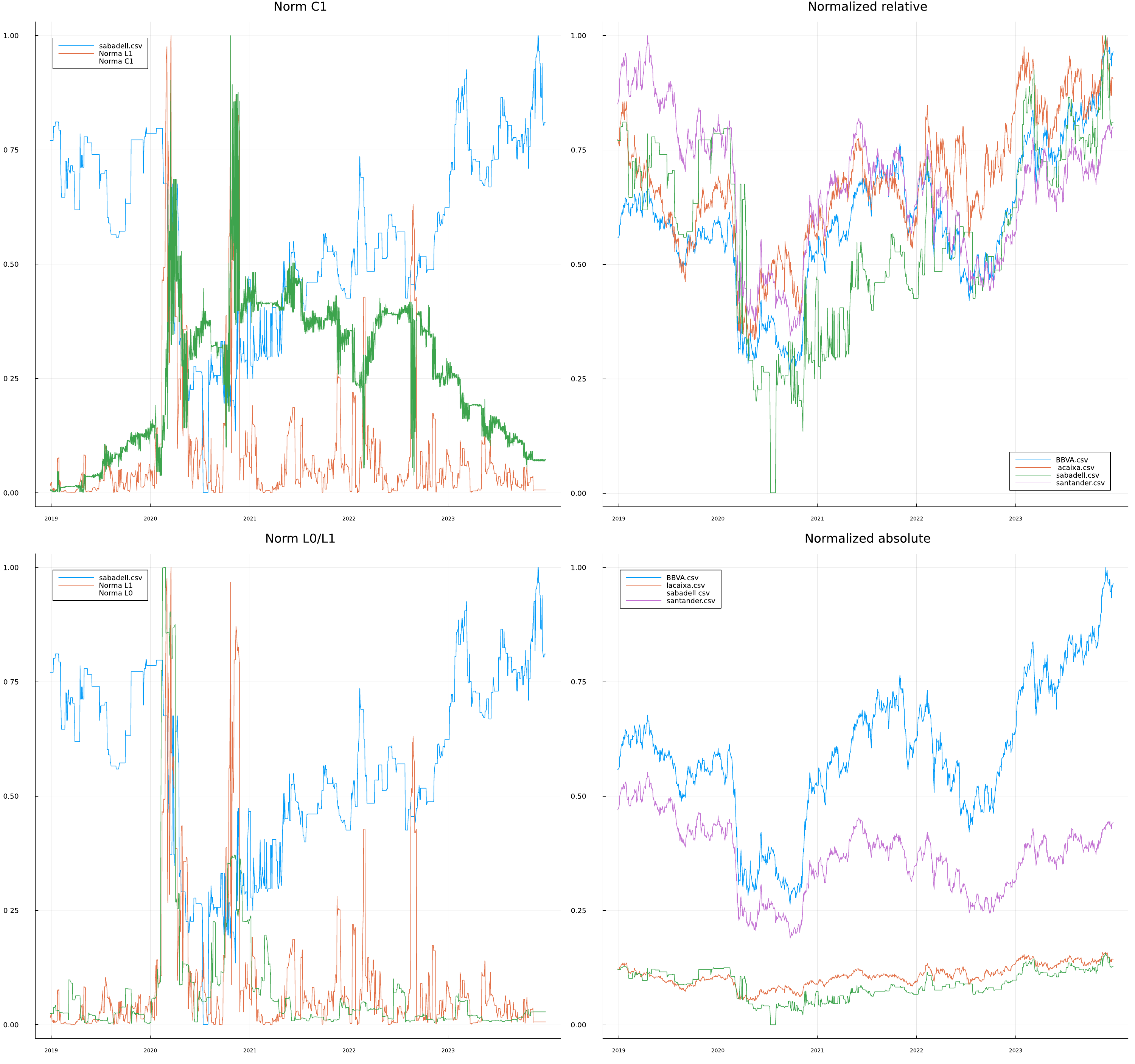}
	\caption{L0,L1,C1 and normalized comparison of Spanish banking indices La caixa / BBVA / Sabadell / Santander}
	\label{fig:banco}
\end{figure}
\newpage
\section{Conclusion}
It is clear that $L1$ norm has great advantages when it comes to time series analysis but it's unable to unravel all the topological information hidden in the point cloud. That's where $L0$ and $C1$ can complement the initial information of $L1$ to determine if a market is stable and whether the market is in a critical transition state. Here we have showcased four examples and how every norm is useful. Upcoming studies of time series event should consider not only the standard $L1$ norm, but also $L0$ to check for outliers and $C1$ for critical transition states.

\begin{landscape}
	
	\begin{figure}[H]
		\centering
		\includegraphics[width=0.8\linewidth]{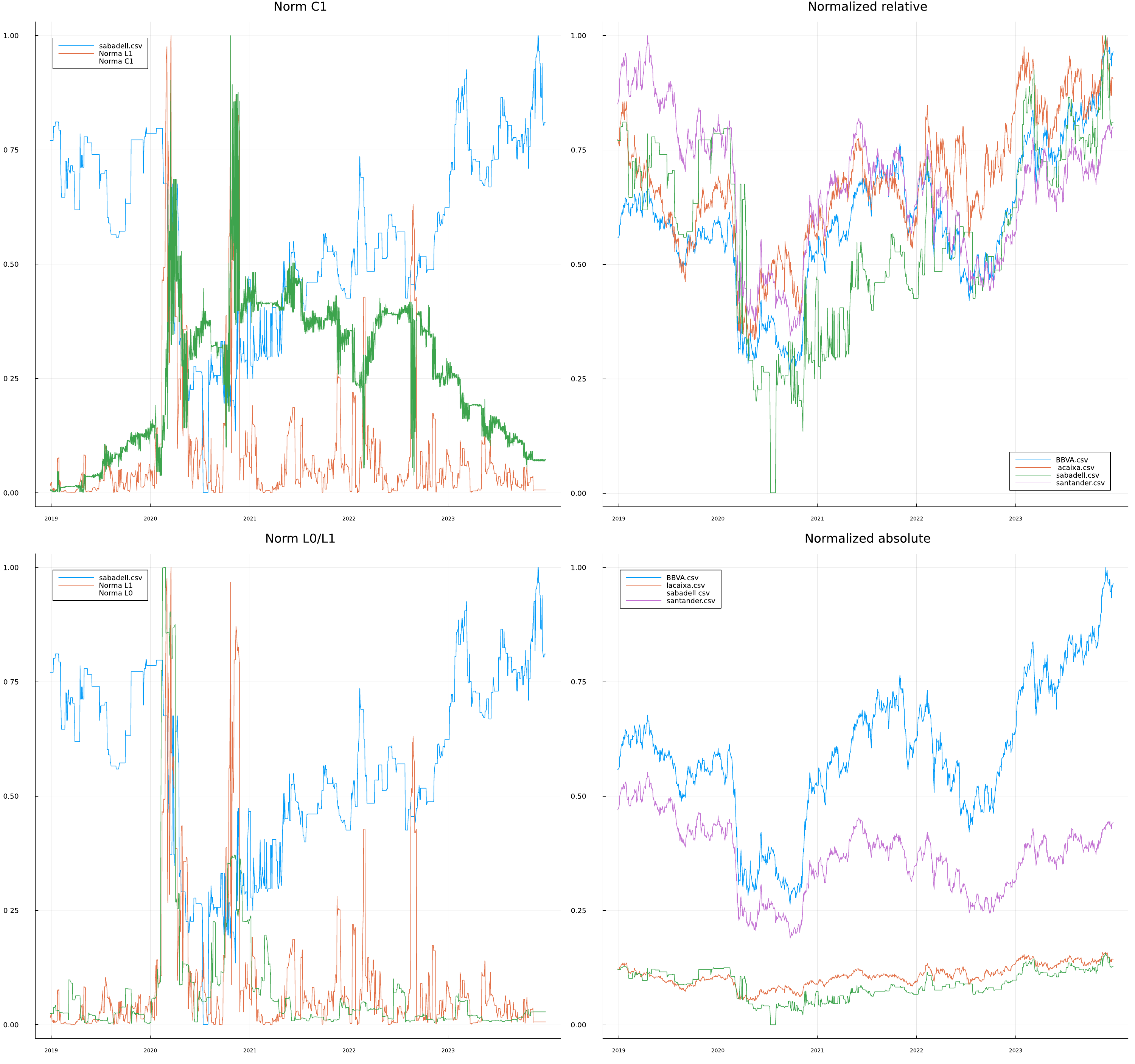}
		\caption{L0,L1,C1 and normalized comparison of Spanish banking indices Lacaixa /BBVA /Sabadell /Santander}
		\label{}
	\end{figure}
	\newpage
\begin{figure}[H]
	\centering
	\includegraphics[width=0.8\linewidth]{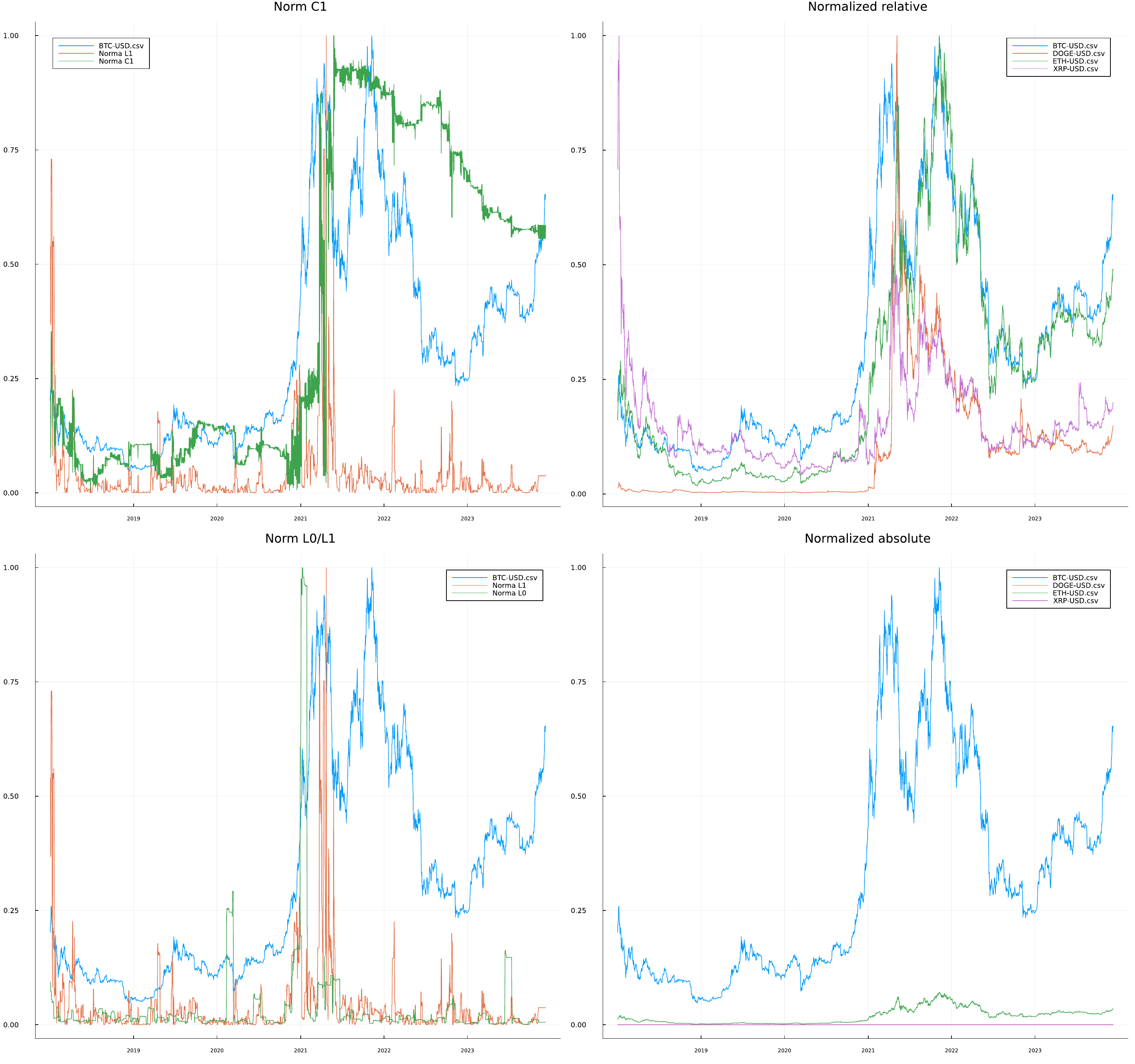}
	\caption{L0,L1,C1 and normalized comparison of cryptocurrency  indices BTC/DOGE/XRP/ETH to USD}
	\label{fig:banco}
\end{figure}
	\newpage
\begin{figure}[H]
	\centering
	\includegraphics[width=0.8\linewidth]{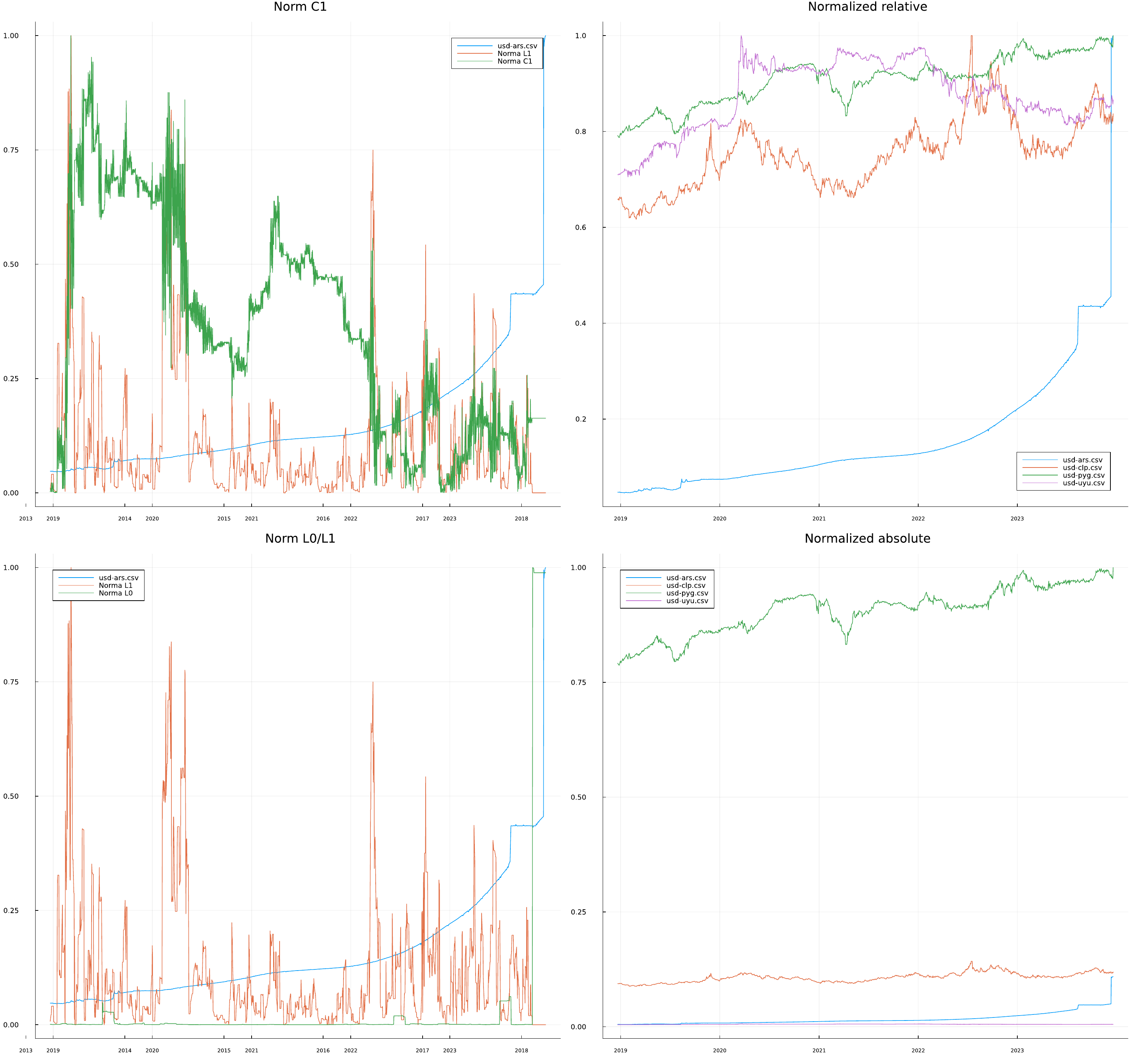}
	\caption{L0,L1,C1 and normalized comparison of South American currencies indices USD to ARS/CLP/PYG/UYU }
	\label{fig:banco}
\end{figure}
	\newpage
\begin{figure}[h]
	\centering
	\includegraphics[width=0.8\linewidth]{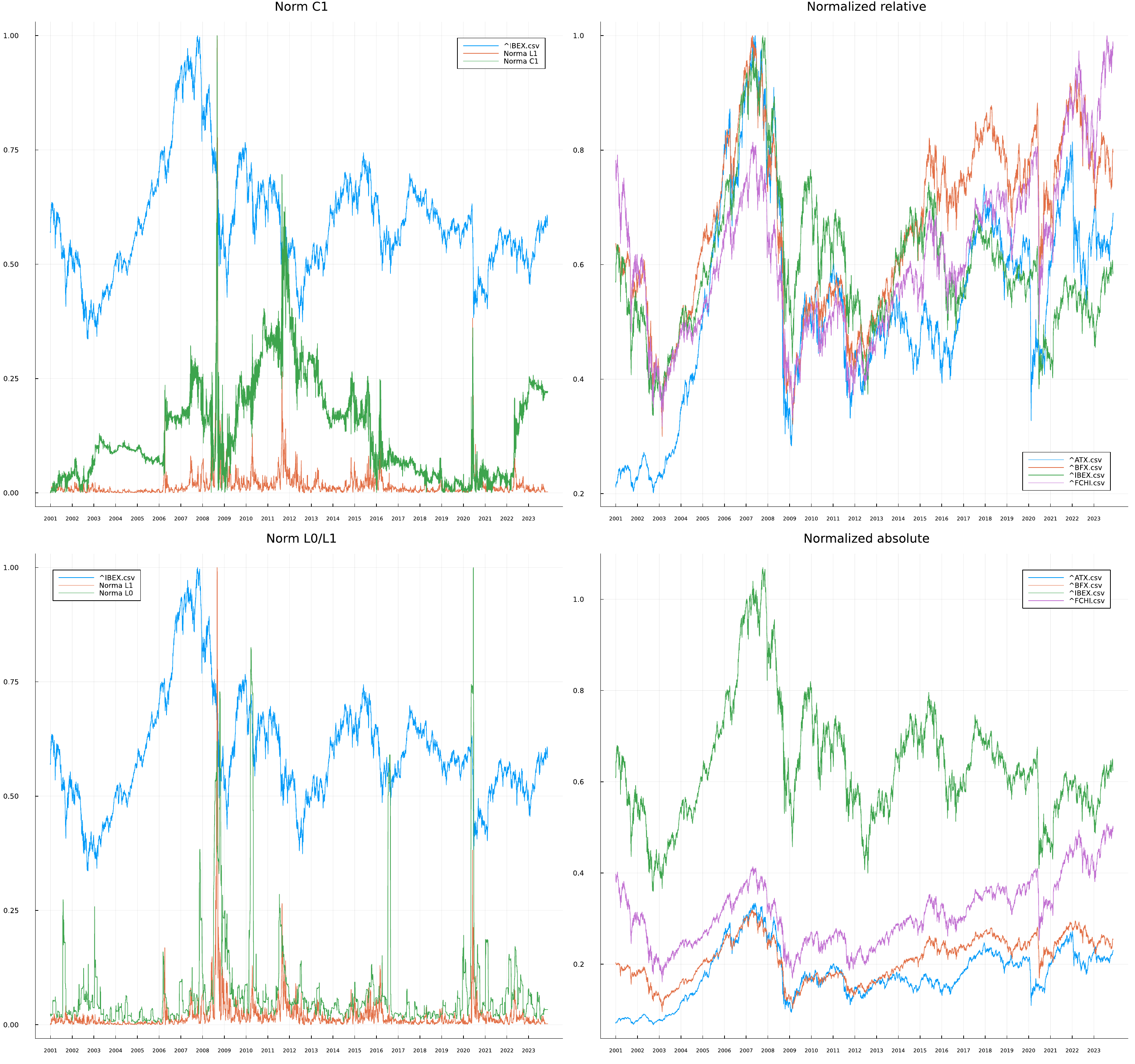}
	\caption{L0,L1,C1 and normalized comparison of european indices IBEX/ATX/BFX/FCHI }
	\label{fig:banco}
\end{figure}
\end{landscape}
\end{document}